\documentclass[fleqn,twoside]{article}
\usepackage{espcrc2}

\usepackage{graphicx}
\usepackage[figuresright]{rotating}

\newcommand{\AmS}{{\protect\the\textfont2
  A\kern-.1667em\lower.5ex\hbox{M}\kern-.125emS}}
\newcommand{\fl}{{\cal F}^I_{J,\lambda}}

\newcommand{\tl}{{\cal T}^I_{J}}
\newcommand{\pppp}{\pi\pi\to\pi\pi}

\newcommand{\kkpp}{{\overline K}K\to\pi\pi}

\title{Illuminating hadron structure by scattering light on light}

\author{M.~R. Pennington\address{IPPP, Physics Department, Durham University, 
        \\ 
        Durham DH1 3LE, U.K.}%
         }
       
\begin{document}

\begin{abstract}
The results of an Amplitude Analysis of the world data on integrated and
differential
cross-sections on $\gamma\gamma\to\pi\pi$ are presented, following the publication of the Belle charged pion results.
\vspace{1pc}
\end{abstract}

\maketitle

\section{INTRODUCTION}

Two photon production of hadronic resonances is one of the clearest ways of revealing their composition, as illustrated in Fig.~1.  The nature of the isoscalar scalars  seen in $\pi\pi$ scattering below 2 GeV, the $f_0(600)$ or $\sigma$, $f_0(980)$, $f_0(1370)$, $f_0(1510)$, $f_0(1720),\, \cdots$,  remains an enigma~\cite{klempt,mp-menu}. While models abound in which some are ${\overline q}q$, some ${\overline {qq}}qq$, sometimes one is a ${\overline K}K$-molecule, and one a glueball, definitive statements are few and far between.
 Since photons couple to neutral hadrons through their charged constituents, as in Fig.~1, their two photon width is a measure of the charge of these~\cite{barnes,barnesKK,achasov,hanhart,menn,mpreviews}. For instance, if the $f_0(980)$ is an ${\overline s}s$ state its radiative width is 0.2 keV~\cite{barnes}, while if it is a ${\overline K}K$-molecule, this is 0.6 keV~\cite{barnesKK} depending on the specifics of the model~\cite{hanhart}. Can experiment distinguish these possibilities?
\begin{figure}[htb]
\begin{center}
\includegraphics[width=10pc]{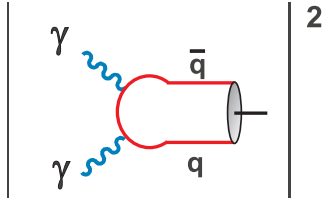}
\caption{Two photon decay rate of a meson in a quark picture is the modulus squared of the amplitude for $\,\gamma\gamma\,$ to produce a $\,{\overline q}q\,$ pair and for these to bind by strong dynamics to form the hadron.}
\end{center}
\vspace{-9pt}
\end{figure}

\begin{figure}[htb]
\begin{center}
\includegraphics[width=16pc]{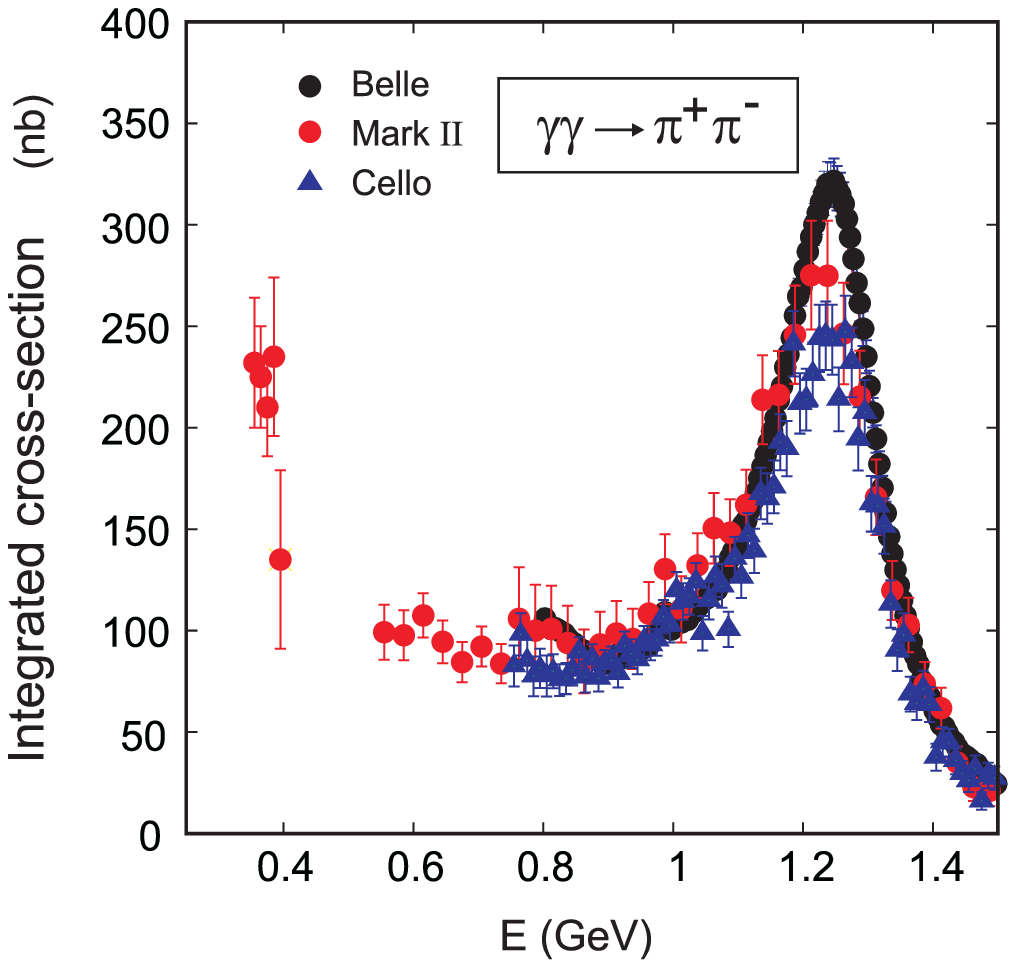}
\vspace{-5mm}
\caption{Comparison of the cross-section results for $\,\gamma\gamma\to\pi^+\pi^-\,$ from Mark~II~\cite{boyer}, Cello~\cite{harjes} and Belle~\cite{abe}. In each case the cross-section is integrated over $|\cos \theta^*|\,\le\,0.6$. $E$ is the $\gamma\gamma$ c.m. energy.}
\end{center}
\vspace{-5mm}
\end{figure}

\begin{figure}[th]
\begin{center}
\includegraphics[width=16pc]{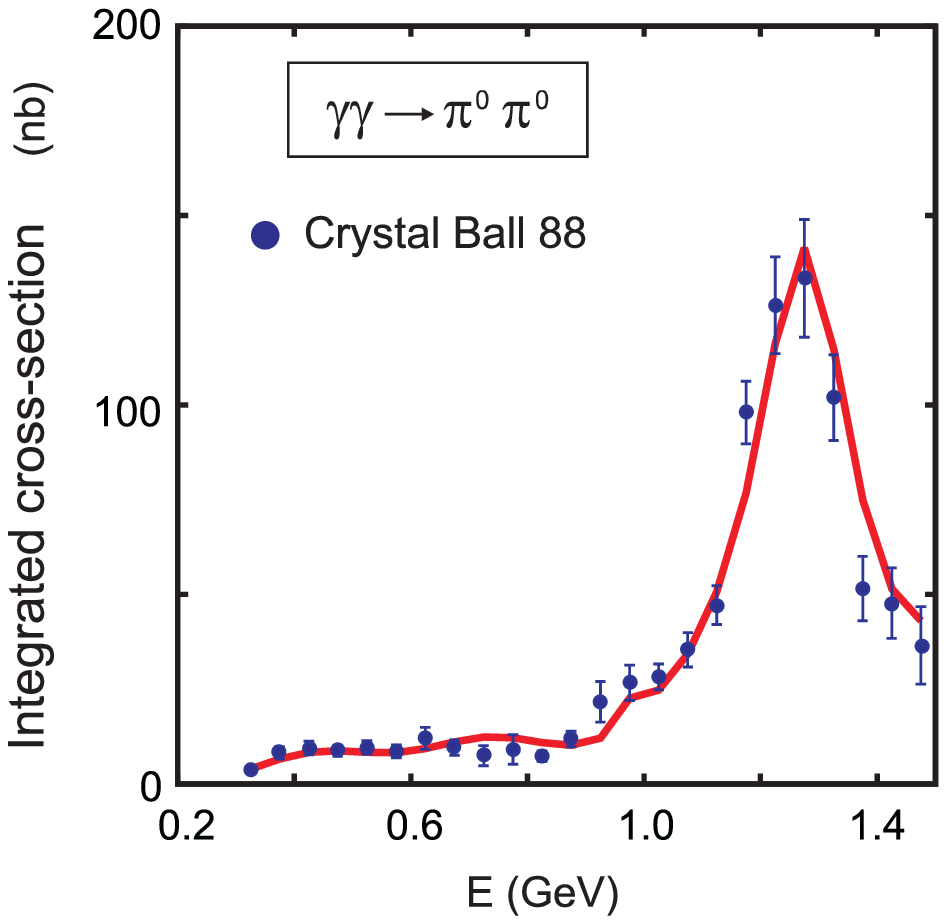}
\vspace{-5mm}
\caption{The favoured Amplitude compared with the Crystal Ball results on 
$\gamma\gamma\to\pi^0\pi^0$. Their 1988 data~\cite{cb88} are integrated over $|\cos \theta^*| \le 0.8$, while the 1992 data~\cite{cb92} (not shown) with increased statistics cover $|\cos \theta^*| \le 0.7$.}
\end{center}
\vspace{-6mm}
\end{figure}
The key features of data on $\pi^+\pi^-$ production~\cite{boyer,harjes,abe}, Fig.~2, are a large enhancement just above threshold, controlled by the one-pion exchange Born term, then a small structure (rather confused in Fig.~2) near 1~GeV  associated with the $f_0(980)$, followed by a clear $f_2(1270)$ peak. The $\pi^0\pi^0$ cross-section, measured (and normalised alone) by Crystal Ball~\cite{cb88,cb92}, Fig.~3, is in contrast small from threshold up to 900 MeV. At 1~GeV, there is small shoulder and then it is dominated by the $f_2(1270)$ signal too. Such dominance reflects the ease with which tensor mesons are formed by two spin-1 photons. But how much of this peak is really pure spin-2? Data on $\gamma\gamma$ production cover only 60-80\% of the angular range, making a complete partial wave separation tricky. However, as we will recall below, by making the most of the general properties of $S$-matrix theory and knowledge of final state hadronic interactions, a determination of the individual spin components becomes feasible. Such a separation of the $\pi^+\pi^-$ and $\pi^0\pi^0$ results published in the 20th century revealed two classes of solutions~\cite{boglione}: one in which the $f_0(980)$ appeared as a {\it peak} with a radiative width of $0.13-0.36$ keV, the other the same state appeared as a {\it dip} with a width of $\sim 0.32$ keV. With data in c.m. energy bins of 20 MeV, both are possible.

The advent of high luminosity $e^+e^-$ colliders with an intense programme of study of  heavy flavour decays has now produced two photon data of unprecedented statistics. The Belle collaboration~\cite{mori,abe} have published results on $\gamma\gamma\to\pi^+\pi^-$ in 5 MeV bins above 800 MeV. These show a very clear peak for the $f_0(980)$, Fig.~4. Analysis of just their integrated cross-section by 
Belle~\cite{abe} finds its radiative width to be $205 ^{+95+147}_{-83-117}$ eV.

\section{AMPLITUDE ANALYSIS}

Here we present the results of an Amplitude Analysis~\cite{belle-mp} of all these data including  the  angular information~\cite{boyer,harjes,behrend,cb88,cb92,abe}. The $\pi\pi$ system can be formed in both $I=0$ and $I=2$ states. The near threshold cross-section is dominated by the Born amplitude, which means that though we expect the $I=2$ $s$-channel to have no resonances, it is comparable to the $I=0$ component in all low energy partial waves.
Consequently we have to treat the $\pi^+\pi^-$ and $\pi^0\pi^0$ channels simultaneously. Though there are now more than 2000 datapoints in the charged channel below 1.5 GeV, we only have 126 in the neutral channel, and we have to weight them more equally to ensure that the isospin components are reliably separable. 

A key role in such an analysis is played by analyticity, unitarity and crossing symmetry.  When these are combined with the low energy theorem for Compton scattering~\cite{low}, and constraints from chiral dynamics, 
\begin{figure}[bh]
\begin{center}
\includegraphics[width=16pc]{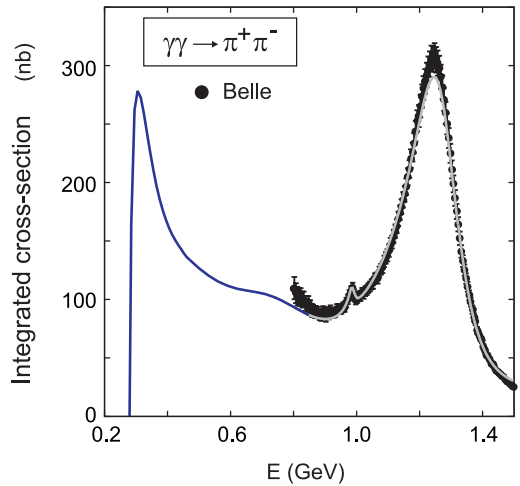}
\vspace{-5mm}
\caption{The favoured Solution compared with the Belle results~\cite{abe}
on $\gamma\gamma\to \pi^+\pi^-$ integrated over $|\cos \theta^*| \le 0.6$.}
\end{center}
\end{figure}
these anchor the partial wave
  amplitudes close to $\pi\pi$ threshold~\cite{morgam}, and help to make up for the limited angular coverage in experiments~\cite{mpdaphne,belle-mp}. Moreover, unitarity imposes a connection between the $\gamma\gamma\to\pi\pi$ partial wave amplitudes and the behaviour of the hadronic processes with $\pi\pi$ final states. As shown in Fig.~5, the relation involves a sum over all kinematically allowed intermediate states \lq$n$' in Fig.~5. 1~GeV marks a divide, below the sum is saturated by the $\pi\pi$ intermediate state, while above the ${\overline K}K$ channel is critically important. Beyond 1.4-1.5 GeV multipion processes start to contribute as $\rho\rho$ threshold is passed. Little is known about the $\pi\pi\to\rho\rho$ channel in each partial wave. Consequently, we restrict attention to the region below 1.44 GeV, where $\pi\pi$ and ${\overline K}K$ intermediate states dominate.
The  hadronic scattering amplitudes, $\tl$, for $\pi\pi\to\pi\pi$ and ${\overline KK}\to\pi\pi$ are known~\cite{bpamps} and so enable the unitarity constraint of Fig.~5 to be realised in practice and in turn allow an Amplitude Analysis to be performed.

\begin{figure}[htb]
\begin{center}
\includegraphics[width=17pc]{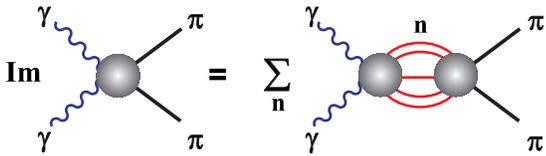}
\vspace{-5mm}
\caption{Unitarity relation for each partial wave of $\gamma\gamma\to\pi\pi$.}
\end{center}
\vspace{-5mm}
\end{figure}
The $\gamma\gamma$ partial waves with definite
isospin $I$, spin $J$ and helicity $\lambda$, ${\cal F}^I_{J,\lambda}(s)$, are parametrised in terms of 
the real functions ${\alpha_i}^{IJ}_\lambda(s)$, where $s$ is the square of the c.m. energy $E$:
\begin{eqnarray}
\nonumber &&\fl(s;\gamma\gamma\to\pi\pi)\;=\;\\[3pt]
\nonumber &&{\hspace{1.2cm}}{\alpha_1}^I_{J\lambda}(s)\,\tl(s;\pppp)\\[3pt]
&&{\hspace{1.8cm}}+{\alpha_2}^I_{J\lambda}(s)\,\tl(s;\kkpp)\, .
\end{eqnarray}
The functions $\alpha(s)$ represent the coupling to each hadronic intermediate state with the appropriate quantum numbers. These couplings, having no right hand cut, are parametrised as smooth functions of $s$, and these form the basis for this energy-dependent Amplitude Analysis.
With data in 5~MeV intervals from Belle, such continuity is sensible. For simplicity,
we will denote the partial waves ${\cal F}^{I=0}_{J,\lambda}$ by $J_\lambda$.

\begin{table*}[tbh]
\caption{Two photon width for the resonances $R$ for the favoured Solution}
\label{table:1}
\vspace{2pt}
\newcommand{\m}{\hphantom{$-$}}
\newcommand{\cc}[1]{\multicolumn{1}{c}{#1}}
\renewcommand{\tabcolsep}{2pc} 
\renewcommand{\arraystretch}{1.4} 
\begin{tabular}{@{}lccc}
\hline
State     & \cc{$f_0(600)/\sigma$} & \cc{$f_0(980)$} & \cc{$f_2(1270)$}\\
\hline
Pole positions (GeV) & $0.441 -i0.272$ & $1.001 - i0.016$ &$1.276 - i0.094$\\
$\Gamma (R\to\gamma\gamma)$ (keV)  & \m$3.1\pm 0.5$ & \m$0.42$ & \m$3.14\pm 0.20$\\
\hline
\end{tabular}\\[5pt]
These widths are determined from the pole residues using Eq.~2. See Ref.~\cite{belle-mp} for the details.
\end{table*}

The world data can be fitted adequately by a range of solutions~\cite{belle-mp}: a range, in which there is a significant ambiguity in the relative amount of helicity zero amplitudes
between $S$ and $D_0$ waves, particularly above 900 MeV. This is a consequence of the data covering only 60-80\% of the angular range,   and hence such waves are not orthogonal to each other in the integrated cross-sections. Nevertheless, there is a favoured solution, which is the one we illustrate here.
  Others are described in the analysis paper~\cite{belle-mp}.

In Figs.~3 and 4, we show how this favoured solution describes the integrated cross-section on $\gamma\gamma\to\pi^0\pi^0$ from Crystal Ball, and on $\gamma\gamma\to\pi^+\pi^-$ from Belle. The resulting amplitude has $I=0$ partial wave cross-sections shown in Fig.~6 for the dominant waves, $S$, $D_0$ and $D_2$ ({\it i.e. $J_\lambda$}). Near threshold their contributions are those of the Born term, modified by largely calculable final state interactions~\cite{morgam}. 
 The $S$-wave shows a clear $f_0(980)$ signal and then a broad enhancement 
\begin{figure}[bh]
\begin{center}
\includegraphics[width=16pc]{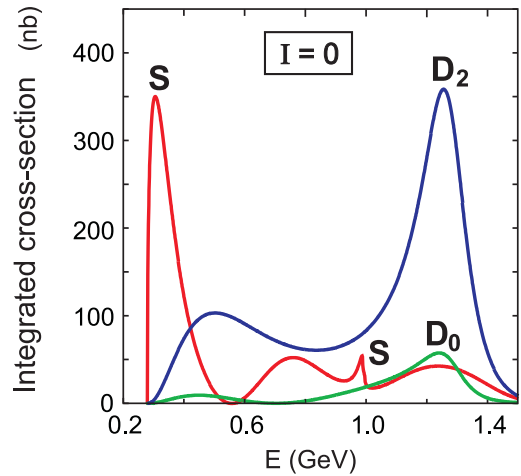}
\vspace{-5mm}
\caption{Contributions of the dominant $I=0$ partial wave components, $J_\lambda$, to the full integrated cross-sections for the favoured Solution.}
\end{center}
\end{figure}
that might be identified with the $f_0(1370)$.
 Above 900 MeV, the $D$ wave is dominated by the $f_2(1270)$, which is produced mainly in the helicity two state, expected for a tensor ${\overline q}q$ meson~\cite{schrempp}.

\section{TWO PHOTON WIDTHS}

In Fig.~7, we show the Argand plot of the favoured $I=J=0$ amplitude, with its clear $f_0(980)$ \lq\lq loop\rq\rq.
By continuing these amplitudes to the resonance pole position on the appropriate unphysical sheet, we determine the two photon coupling of the $f_0(980)$ and for the $D$-wave the $f_2(1270)$.
These pole residues are the only process-independent definition of their couplings, $g_\gamma$.
 These give a measure of their radiative widths through the commonly used formula:

\begin{equation}
\Gamma(R\to\gamma\gamma)\;=\;\frac {\alpha^2}{4 (2J+1) m_R}\;| g_{\gamma}|^2 \quad ,
\end{equation}
where $\alpha \sim 1/137$ is the fine structure constant.
This gives the values listed in Table 1.
As described in Ref.~\cite{belle-mp}, the full range of acceptable solutions gives a radiative width for the $f_0(980)$ from 100 to 540 eV, with 415 eV favoured. This range allows ${\overline s}s$, ${\overline K}K$ and ${\overline{qq}}qq$ compositions to be possible~\cite{barnes,barnesKK,achasov,hanhart,menn,mpreviews}. However, the favoured solution accords with none of these specific alternatives.

\begin{figure}[t]
\begin{center}
\includegraphics[width=16.pc]{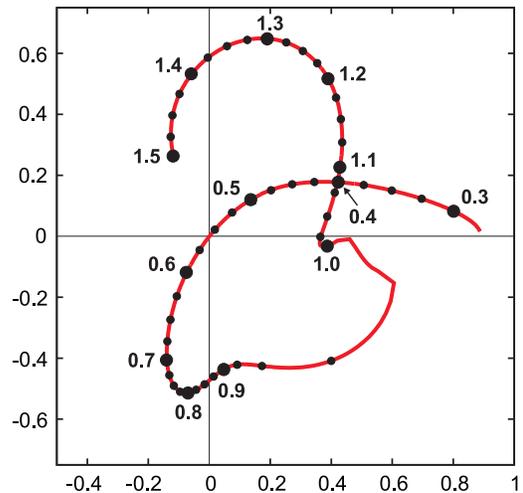}
\vspace{-4mm}
\caption{Argand plot for the favoured $\gamma\gamma\to\pi\pi$ $I=0$ $S$-wave amplitude. 
 The labels mark the energy every 0.1 GeV, 
with smaller dots every 25 MeV. 
The amplitude moves quickly between 950 and 1000 MeV 
because of the $f_0(980)$, with  \lq\lq kinks'' 
at the two ${\overline K}K$ thresholds.
}
\end{center}
\vspace{-3mm}
\end{figure}
Achasov and Shestakov~\cite{achasov-belle} have analysed the Belle integrated
cross-sections in terms of amplitudes in which key resonances have both direct and meson loop contributions, and they find
$\Gamma(f_0\to\gamma\gamma) \simeq 0.2$ keV, which supports their 
model of the $f_0(980)$ as a  ${\overline {qq}}qq$  state. 

Here we have presented the amplitude favoured by world data on  both 
integrated and differential cross-sections for both charged and neutral pion production. As discussed in more detail in Ref.~\cite{belle-mp}, this is one of a range of amplitudes that provide an acceptable description of current experiments. The forthcoming $\gamma\gamma\to\pi^0\pi^0$ data from Belle~\cite{nak} should have the power to reduce this range considerably. Once these are finalised the inclusion of these data in an Amplitude Analysis will hopefully lead to a consistent set of two photon widths for the low mass isoscalar states.   

For the $f_0(600)$ or $\sigma$ in Table 1, we have little new to add~\cite{mp-prl,oller} since the Belle data only start at 800 MeV, Figs.~2,4, and so the analysis relies on older data~\cite{boyer,cb88} --- see Ref.~\cite{belle-mp} for the discussion. To reduce still further the uncertainty in its $\gamma\gamma$ width requires precision charged and neutral pion data between threshold and 700~MeV~\cite{mpreviews}. The introduction of appropriate taggers
in an upgraded DA$\Phi$NE machine at Frascati~\cite{daphne2} may well make this feasible.

 Two photon couplings are a key window on the detailed structure of low mass 
scalar states. Such couplings are likely to be just as critical in  determining the nature of the scalar field(s) that break electroweak symmetry. Perhaps this/these scalar(s) that await discovery in the $0.1-1$ TeV region will be just as complex in their structure as those at $0.1-1$ GeV, making the present study of the states of QCD an important guide. 

\vspace {5.5mm}

The author acknowledges partial support of the EU-RTN Programme, Contract No. MRTN--CT-2006-035482, \lq\lq Flavianet'' for this work.

\vspace{0.5mm}


\end{document}